# Discovering like a scientist: Literature-informed dimensionless learning for physical closure of data-driven equations

Hao Xu[1], Tianhao Wu[1,*], and Dongxiao Zhang[1,*]

[1] Zhejiang Key Laboratory of Industrial Intelligence and Digital Twin, Eastern Institute of Technology, Ningbo, Zhejiang 315200, China.

[*] Corresponding authors.

Email address: twu@eitech.edu.cn (T. Wu); dzhang@eitech.edu.cn (D. Zhang).

**Abstract**

Data-driven methods can recover compact equations, yet the physical dependence of their coefficients across operating conditions often remains unresolved. Dimensionless groups can explain this variation, but selecting relevant groups from the many dimensionally admissible combinations remains challenging. Here, we introduce a knowledge-guided framework for dimensionless learning, called π-former. It learns patterns of variable association and exponent assignment from a corpus of 1,288 established dimensionless groups. Under exact dimensional constraints, the model proposes interpretable basic groups whose combinations are optimized against observations. Across seven physical problems, the framework identifies predictive dimensionless coordinates within seconds. Integration with equation discovery further expresses condition-dependent coefficients through these coordinates, recovering established physical scalings in multi-condition simulations. Applied to experimental measurements of evaporation-driven wetting, the framework discovers dimensionless coordinates and uses them to construct a unified trajectory equation. The resulting model generalizes to held-out conditions across liquids and initial volumes, outperforming existing empirical models. These results establish dimensionless learning as a transferable bridge between accumulated scientific knowledge and data-driven equation discovery, enabling artificial intelligence to uncover physically meaningful laws from observations.

## Introduction

Data-driven methods can now recover compact equations directly from observations. Symbolic regression has distilled free-form relations from experimental measurements[1,2] , yielding constitutive laws[3] and algebraic models[4]. Sparse identification[5,6] and neural approaches[7–9] have extended equation discovery to nonlinear dynamical systems, enabling the identification of partial differential equations governing diverse physical processes[10,11]. Yet the coefficients of equations recovered from individual datasets can be difficult to interpret physically and often vary across operating conditions. Establishing a transferable physical law therefore requires identifying physical coordinates that explain this variation through a unified relation.

Dimensionless numbers provide such coordinates by expressing balances among physical mechanisms independently of the choice of units and defining similarity across scales[12] . The Reynolds number, for example, compares inertial and viscous effects[13]. Geometrically similar flows at the same Reynolds number share this balance despite differences in size and speed[14]. Dimensionless coordinates can therefore reduce the effective degrees of freedom, collapse measurements from different conditions and connect laboratory observations to full-scale systems. Buckingham's Pi theorem formalizes their construction by placing admissible monomials in the null space of the dimensional matrix[15]. The theorem determines the number of independent groups while leaving the choice of basis open. Infinitely many equivalent representations can consequently be formed, and the search remains large even when exponents are restricted to a finite range. Scientific practice resolves this ambiguity through mechanistic reasoning and empirical tests of similarity. The established dimensionless groups collected across scientific fields preserve the outcomes of this process.

Machine learning has begun to automate the selection of dimensionless coordinates. Active-subspace formulations identify dimensionless ridge coordinates that capture the dominant variation of a response surface[16] . BuckiNet places a dimensionally constrained projection before the predictor and learns the associated exponents by minimizing prediction error together with dimensional-consistency and regularization penalties[17]. Dimensionless learning extends this strategy through bilevel optimization tied to predictive fit or data collapse[18]. The recent IT-π framework replaces dependence on a chosen predictor with an information-theoretic measure of irreducible error, which supports model-independent ranking and regime analysis[19]. These studies show that observations can distinguish useful coordinates among dimensionally admissible candidates. However, each new variable set is generally treated as an independent search problem, while construction patterns encoded in established dimensionless groups remain outside the learning objective. Several frameworks have also used discovered dimensionless groups to parameterize coefficients in sparse equations[17,18]. Yet applying dimensionless learning to achieve physical closure in real experimental systems remains underexplored. Such closure is important for scientific practice because it connects discovered equations to underlying physical conditions and supports transfer across experiments.

Here we develop π-former, a knowledge-guided framework that learns recurring patterns of variable association and exponent assignment from established dimensionless groups. It proposes basic groups under exact Buckingham constraints, allowing observations to guide an optimization over a compact set of learned hypotheses. π-former reconstructs established groups in held-out corpus examples and identifies the best-performing coordinates across seven physical problems within seconds. It also explains condition-dependent coefficients in equations discovered from multiple datasets, recovering the expected Reynolds dependence in cavity flow and the Damköhler–Péclet dependence in reaction–advection–diffusion dynamics. Applied to experimental measurements of evaporation-driven wetting in a thin porous medium, π-former identifies dimensionless coordinates for the maximum spreading radius and characteristic time. Using these scales, graph-based equation discovery yields a compact trajectory relation that satisfies the physical condition at maximum spreading and transfers to held-out experimental conditions and outperforms existing empirical models.

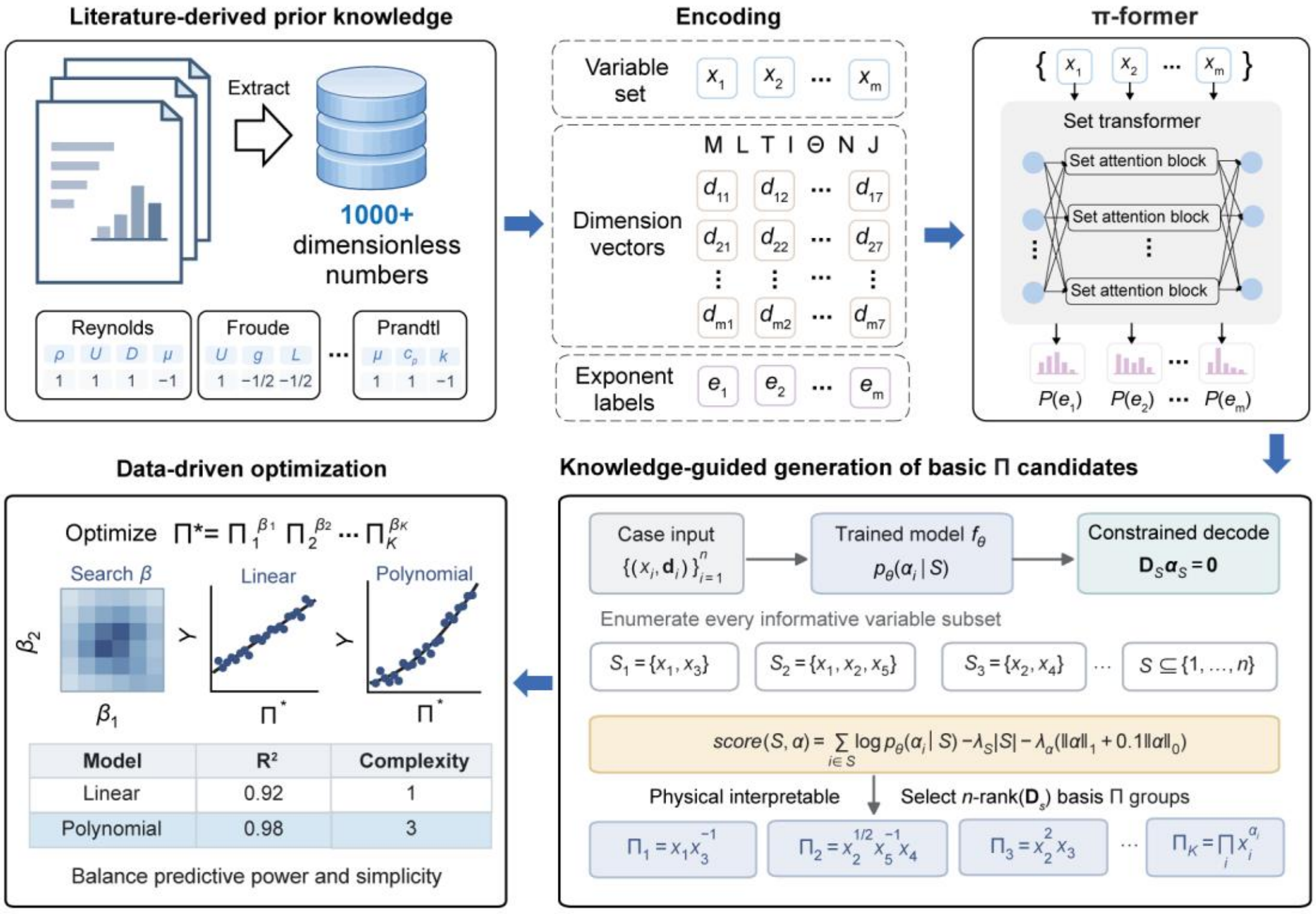

**Fig. 1. Overview of the proposed framework to learn dimensionless groups from scientific literature.** Existing dimensionless numbers recorded in the literature are encoded as unordered variable sets with dimension vectors and exponent labels. A set transformer (π-former) learns the mapping from each variable set to corresponding exponents. On a new case, the trained π-former generates basic dimensionless groups (Π) on informative variable subsets, subject to the homogeneity constraint $D_S \alpha_S = 0$, and ranks them by a penalized likelihood. Data-driven optimization then forms a composite coordinate by optimizing through the observation data.

## Results

### Generation of dimensionless groups with π-former

As summarized in Fig. 1, we constructed a corpus of 1,288 dimensionless groups automatically extracted from Kunes's handbook[20], with human review and dimensional-homogeneity checks ensuring internal consistency. Using this corpus, we train π-former, a Set Transformer[21], to predict exponent distributions for each input variable. Its permutation-equivariant architecture preserves each variable's prediction under input reordering. By learning patterns of variable association and exponent assignment, π-former generates hypotheses for basic dimensionless groups. More details of π-former are provided in the Methods. For a new problem, π-former encodes variable dimensions and proposes exponent assignments under Buckingham's π constraint. It evaluates all admissible variable subsets, allowing basic dimensionless groups to exclude redundant quantities. The highest-scoring candidates are retained as basic groups (Fig. 1) and combined through a product of powers, with composition exponents optimized against observations. This knowledge-guided approach narrows the search for problem-specific dimensionless coordinates while preserving recognizable physical structure.

We first evaluated the generalization of π-former through ten-fold cross-validation, using 80% of the corpus for training, 10% for validation, and 10% for testing in each fold. π-former reconstructed held-out dimensionless groups from their variables and dimensional encodings through constrained decoding, ranking admissible exponent vectors by learned scores. π-former achieves high recovery rates across variable-set sizes (Fig. 2a). Top-1 recovery exceeds 90% for groups with three variables or fewer, while Top-5 recovery remains above 80% for four to six variables and above 60% even for the more challenging groups with seven or more variables. Overall, the recorded dimensionless group ranks first in 82.3% of held-out cases and within the Top-5 in 94.0% (Fig. 2b). These results show that π-former concentrates physically plausible dimensionless groups within a short candidate list before data-driven optimization, narrowing the subsequent search.

Fig. 2c compares the rank-1 and rank-2 hypotheses with the empirical groups for two examples. For the condensation number $K_\delta$, which was absent from training, π-former uses only the variable set and the dimensional matrix and returns structures close to the empirical form. The rank-1 candidate differs from the recorded group only by an interchange of the two density terms, whereas the rank-2 candidate is identical to the recorded group. Both candidates are dimensionally admissible, and their distinction should therefore be determined by data-driven fitting. The specific-diameter example shows that the same procedure tracks fractional exponents. π-former draws exponents from a discrete set of rationals that recur in the empirical corpus, rather than from a continuous exponent space. The restriction keeps the proposed exponents in the simple rational forms that experts ordinarily write, and it concentrates the search on groups that remain interpretable.

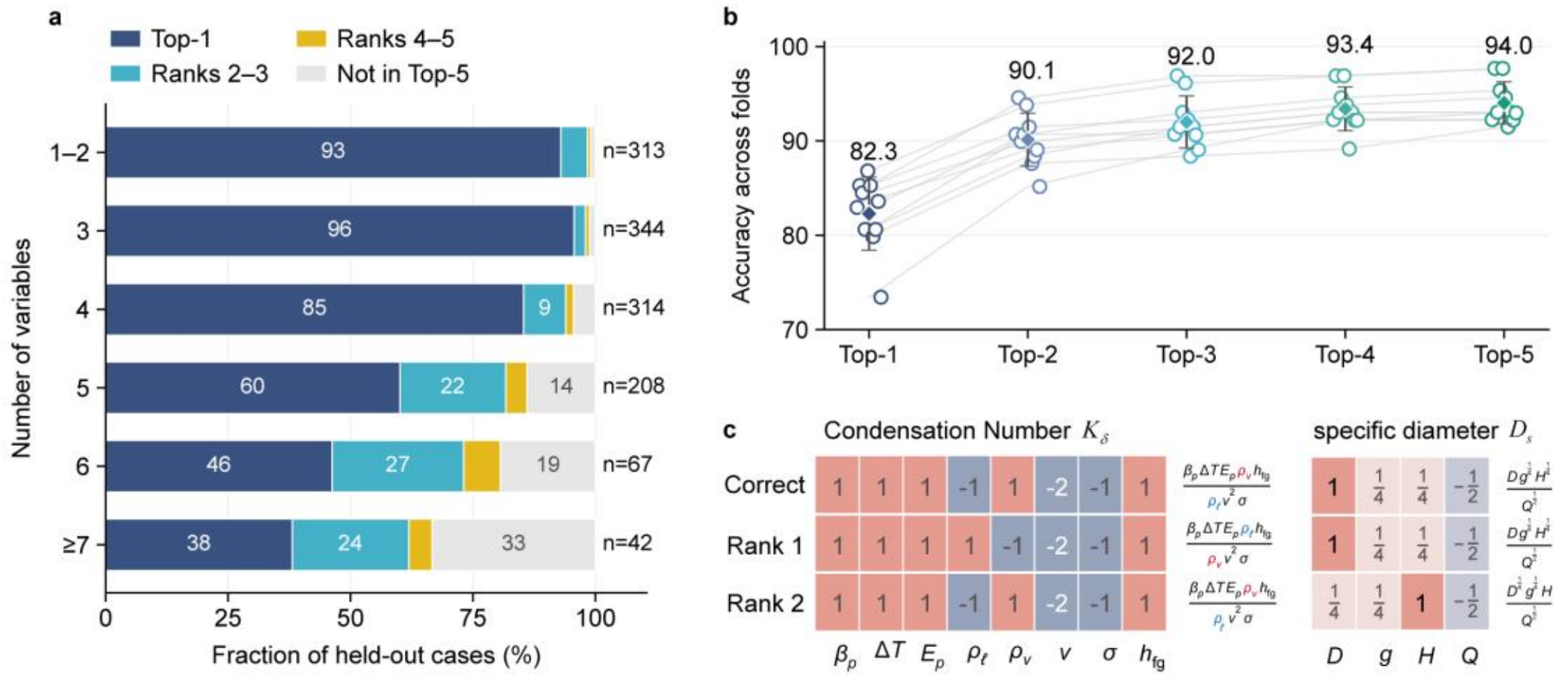

**Fig. 2. Held-out recovery of literature dimensionless numbers by π-former. (a)**, Rank of the true exponent pattern among model-generated candidates, stratified by the number of variables in the group. $n$ is the number of cases in each bin. **(b)**, Top-$k$ accuracy across cross-validation folds. Circles are individual folds. **(c)**, Ground-truth exponents and the top two predicted dimensionless groups for the condensation number and the specific diameter.

**Recovery of canonical dimensionless coordinates**

We next tested whether the proposed framework can combine π-former hypotheses with observations and recover textbook coordinates from raw physical variables. The first problem was Rayleigh-Bénard convection, in which heating from below drives buoyancy-induced fluid motion and vertical heat transport[22]. The problem involved seven variables. Its dimensional matrix has rank four and therefore admitted three independent dimensionless groups (Fig. 3a). We accordingly retained the three highest-ranked hypotheses from π-former as the basic dimensionless groups:

$$\Pi_1 = \alpha \Delta T, \qquad \Pi_2 = \frac{g\alpha \Delta T h^3}{\nu\kappa}, \qquad \Pi_3 = \frac{\nu}{\kappa}. \tag{1}$$

Each dimensionless group has a direct physical interpretation. $\Pi_1$ measures the relative density variation induced by thermal expansion. $\Pi_2$ and $\Pi_3$ coincide with the Rayleigh and Prandtl numbers. To keep the test uncontaminated, all dimensionless numbers associated with this process were removed from the corpus and π-former was retrained.

A power product of the three groups defined a coordinate for predicting the Nusselt number. Starting with linear regression on a single coordinate, the search increased regression order before adding coordinates as needed to meet the target accuracy. The exponent vector $\beta$ was fitted using 80% of the observations, with 20% held out for testing, and candidates were ranked and selected by the coefficient of determination ($R^2$). For Rayleigh-Bénard convection, the optimized exponent vector was $\beta^* = [0,1,0]^{\mathrm{T}}$, which retains only $\Pi_2$ (Fig. 3b). Using a polynomial fit, the discovered coordinate accounts for the Nusselt number with $R^2 = 0.999$ and a mean absolute percentage error of 5.1% (Fig. 3c). π-former ranks the Prandtl number among its leading hypotheses despite its near constancy in the Rayleigh–Bénard data and absence from the fitted law. It thus proposes physically relevant groups beyond those selected by data-driven fitting.

The second problem concerned the Colebrook relation for turbulent flow through a rough pipe, in which the Darcy friction factor depends on the Reynolds number and relative wall roughness[23]. This case involved five variables. The dimensional matrix had rank three and admitted two independent coordinates (Fig. 3d). After applying the same exclusion and retraining procedure, π-former generated:

$$\Pi_1 = \frac{u\rho D}{\mu}, \qquad \Pi_2 = \frac{u\rho k}{\mu} \tag{2}$$

as the two leading basic dimensionless groups. Both have the form of a Reynolds number. The first uses the pipe diameter as the characteristic length, whereas the second uses the roughness height. The workflow initially searched for a single composite coordinate. Optimization produced $\boldsymbol{\beta}=[1,-1]^T$, giving

$$\Pi^* = \frac{\Pi_1}{\Pi_2} = \frac{D}{k} \tag{3}$$

which is the inverse relative roughness. This coordinate achieved a test $R^2$ of 0.794 (Fig. 3e). This triggered the search for a two-coordinate representation. The best model used $\Pi_1$ together with $(\Pi^*)^{-1}=k/D$, corresponding to the diameter-based Reynolds number and relative roughness. Their joint representation increased the test $R^2$ to 0.887 with linear fitting (Fig. 3f), and the resulting surface captured the observed dependence of the friction factor across the two-dimensional coordinate space (Fig. 3g).

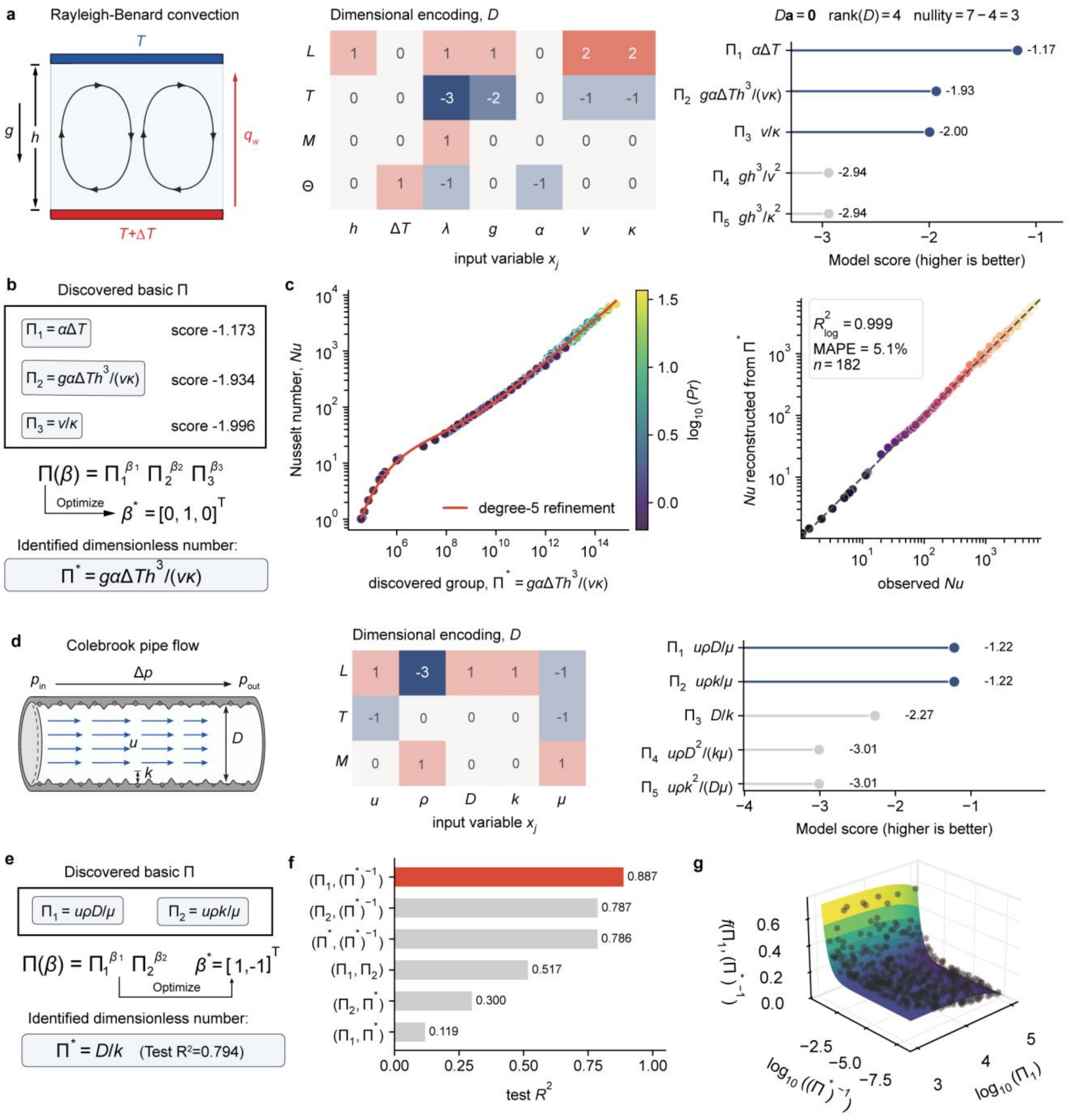

**Fig. 3. Recovery of canonical dimensionless numbers. (a),** Physical configuration, dimensional encoding and ranked candidate groups for Rayleigh-Bénard convection. **(b),** Selected basic groups and data-driven composition yielding the Rayleigh number. **(c)**, Polynomial relation between the discovered dimensionless number and the Nusselt number (left), and reconstructed versus observed Nusselt numbers (right). **(d),** Physical configuration, dimensional encoding and ranked candidate groups for Colebrook pipe flow. **(e)**, Generated basic groups and optimized dimensionless number. **(f),** Test $R^2$ values for candidate combination of two dimensionless numbers. **(g),** Fitted surface over the selected Reynolds-number and relative-roughness coordinates.

## Comparison with existing methods for dimensionless learning

To clarify the differences between the proposed procedure and existing approaches, we compared it with three representative dimensionless-number discovery methods: BuckiNet[17], IT-π[19] and PyDimension[18]. Each baseline used the case-specific settings provided in its public implementation. Of note, BuckiNet follows the neural-network path and imposes the Buckingham condition as a soft penalty on the null space[17]. The comparison covered seven physical problems. Four were canonical cases with established empirical dimensionless groups. The remaining three were open problems for which suitable dimensionless numbers had to be inferred from data. Details of these problems are provided in Supplementary Information S.1. We recorded the highest-performing dimensionless group and runtime for each method. To ensure a fair comparison of computational cost, all methods were run on the same CPU. Computations were performed serially, with GPU acceleration disabled. For data-driven optimization, each case used 80% of all available data for training and the remaining 20% for testing. Details of the dataset are provided in Table S1. The data split and random seed were held fixed across methods. The results are summarized in Table 1.

The proposed method achieved the shortest runtime in most cases and typically completed the search within approximately one second. The reduction in runtime became more pronounced as dataset size increased. Conventional methods rely on computationally intensive continuous optimization. For example, BuckiNet trains a neural network to optimize the exponents, whereas IT-$\pi$ uses CMA-ES to search for coefficients over a dimensional basis. GPU acceleration or CPU parallelism can bring a single run down to a few seconds or a few tens of seconds, yet the computational budget of both searches remains large. $\pi$-former instead generates physically interpretable hypotheses informed by exponent patterns in the scientific literature, reducing the subsequent search to a discrete set of common rational values. This design assumes that basic dimensionless groups should retain simple exponent structures, whereas complex scaling laws can be fitted in logarithmic space. For example, a response may scale with the Reynolds number as a non-integer power, while the Reynolds number itself retains its standard form. The resulting groups therefore combine concise exponent structures with comparable held-out $R^2$, balancing simplicity and predictive accuracy.

For all four canonical problems, the proposed framework recovered the established dimensionless forms from the given variables and observations. In contrast, BuckiNet and PyDimension failed to recover the established dimensionless groups in several cases, particularly for problems involving multiple groups. For the open problems, the proposed method identified dimensionless coordinates with predictive performance that was best or closely competitive across the comparison. The resulting groups also retained simple exponent structures drawn from the learned construction prior. In the Supplementary Information S.2 and Fig. S1, we examined how the candidate set for exponent $\beta$ affects computational cost. Even the densest set required only about one minute per optimization, while the expanded search space further improved accuracy for complex cases such as Keyhole. Notably, $\pi$-former not only returns the final optimized dimensionless group, but also provides a set of physically meaningful basic dimensionless groups (Table S2). Although some of them may be dropped during data-driven optimization, these candidates reveal additional physical relationships that aid interpretation of the system. By comparison, existing methods derive null-space basis vectors through SVD or require the basis to be specified manually.

**Table 1. Recovered dimensionless numbers and computational time for seven problems with four different methods.** For the three open problems without empirical dimensionless numbers, including MHD generator, Dixit and Keyhole, the test $R^2$ is also reported.

| Case | BuckiNet | IT-π | PyDimension | Ours |
|---|---|---|---|---|
| **Rayleigh–Bénard** | $\frac{\lambda^{0.746}\ g\ \kappa^{0.352}}{h^{0.871}\ \Delta T^{0.24}\alpha^{0.487}\ \nu^{0.09}}$<br>$t$=2.01 s | $\frac{h^3(\Delta T)^{1.08}g^{0.99}\alpha^{1.08}}{\nu^{0.99}\kappa^{1.02}}$<br>$t$=4.1 s | $\frac{h^3(\Delta T)g\alpha}{\nu\kappa}$<br>$t$=0.336 s | $\frac{h^3(\Delta T)g\alpha}{\nu\kappa}$<br>$t$=0.26 s |
| **Colebrook** | $\Pi_1 = uD^{0.985}$<br>$\Pi_2 = \frac{D^{0.998}}{k}$<br>$t$=379 $s$ | $\Pi_1 = \frac{k^{0.99}}{D}$<br>$\Pi_2 = \frac{\mu}{u\rho D}$<br>$t$=690.6 $s$ | $\Pi_1 = \frac{k^{0.875}\mu^{0.125}}{u^{0.125}\rho^{0.125}D}$<br>$\Pi_2 = \frac{u\rho D^{0.37}k^{0.63}}{\mu}$<br>$t$=6.29 $s$ | $\Pi_1 = \frac{u\rho D}{\mu}$<br>$\Pi_2 = \frac{k}{D}$<br>$t$=1.52 $s$ |
| **Blasius** | $\frac{Uy^{0.268}}{\nu^{0.37}x^{0.129}}$<br>$t$=84.4 $s$ | $\frac{U^{0.51}y}{\nu^{0.51}x^{0.49}}$<br>$t$ = 26.6 $s$ | $\frac{U^{0.618}y}{\nu^{0.618}x^{0.382}}$<br>$t$ = 0.508 $s$ | $\frac{U^{1/2}y}{\nu^{1/2}x^{1/2}}$<br>$t$=0.182 $s$ |
| **Rayleigh** | $\frac{y}{t^{0.5}\nu^{0.5}}$<br>$t$=497 $s$ | $\frac{t^{0.5}\nu^{0.5}}{y}$<br>$t$=149.2 $s$ | $\frac{t^{0.388}\nu^{0.612}}{U^{0.225}y}$<br>$t$=3.81 $s$ | $\frac{y^2}{t\nu}$<br>$t$=1.74 $s$ |
| **MHD generator** | $\frac{\rho\left(\frac{\mathrm{d}p}{\mathrm{d}x}\right)^{0.0252}}{\mu^{0.712}}$<br>$R^2$= 0.609<br>$t$=5.12 $s$ | $\frac{\ell\rho^{0.33}\left(\frac{\mathrm{d}p}{\mathrm{d}x}\right)^{0.33}}{\mu^{0.67}}$<br>$R^2$= 0.987<br>$t$=11.0 $s$ | $\frac{\ell\rho^{0.326}\left(\frac{\mathrm{d}p}{\mathrm{d}x}\right)^{0.326}}{\mu^{0.663}}$<br>$R^2$=0.987<br>$t$=0.144 $s$ | $\frac{\ell^3\rho\left(\frac{\mathrm{d}p}{\mathrm{d}x}\right)}{\mu^2}$<br>$R^2$=0.998<br>$t$=0.043 $s$ |
| **Dixit** | $\frac{U_\infty^{0.0827}(\delta^*)^{0.553}\theta}{\nu^{0.613}M^{0.5}}$<br>$R^2$=0.882<br>$t$=0.877 $s$ | $\frac{U_\infty\delta^{0.55}(\delta^*)^{0.69}}{M^{0.44}\theta^{0.68}\nu^{0.12}}$<br>$R^2$=0.954<br>$t$=5.8 $s$ | $\frac{\nu^{0.18}M^{0.41}\theta^{0.856}}{U_\infty\delta^{0.46}(\delta^*)^{0.986}}$<br>$R^2$= 0.926<br>$t$=1.77 $s$ | $\frac{U_\infty^{5/2}\delta\delta^*}{\nu^{1/2}M\theta^{1/2}}$<br>$R^2$= 0.948<br>$t$=1.06 $s$ |
| **Keyhole** | $\frac{\eta P\rho^{0.253}c_p^{0.532}\alpha^{0.955}}{V_s^{0.192}(\Delta T)^{0.335}r_0^{0.481}}$<br>$R^2$=0.650<br>$t$ = 1.71 $s$ | $\frac{\eta P^{0.6}}{V_s^{0.3}\rho^{0.7}C_p^{0.7}(\Delta T)^{0.7}r_0\alpha^{0.4}}$<br>$R^2$ = 0.980<br>$t$ = 2.3 $s$ | $\frac{\eta P}{\Delta T\pi\rho C_p\sqrt{\alpha V_s r_0{}^3}}$<br>$R^2$ = 0.978<br>$t$ = 0.515 $s$ | $\frac{V_s^{1/2}\rho c_p\Delta T r_0^{3/2}\alpha^{1/2}}{\eta P}$<br>$R^2$ = 0.978<br>$t$ = 0.368 $s$ |

**Discovering dimensionless closure of condition-dependent governing laws**

We next integrated dimensionless learning with partial differential equation discovery to construct an end-to-end process from field observations to condition-dependent governing laws. For each system, numerical simulations were generated under multiple physical settings, producing distinct field evolutions across cases. A genetic algorithm constrained by physical symmetry first searched for a differential structure shared among these cases[24]. More details about the method can be found in Supplementary Information S.3. Once the structure was determined, its coefficients were estimated separately under each condition. Their variation was then treated as an unresolved physical dependence. Dimensionless learning supplied potential dimensionless numbers for this dependence and expressed the varying coefficients as explicit closure relations.

The first example considered vorticity fields from ten lid-driven cavity-flow cases[25] spanning Reynolds numbers from 50 to 1,000 (Fig. 4a). The cases used different values of density, velocity, length and viscosity, which produced visibly different flow fields. Independent genetic searches discovered the most frequently recovered structure

$$\omega_t = c_0 u\omega_x + c_1 v\omega_y + c_2\omega_{xx} + c_3\omega_{yy} \quad (4)$$

in six of the ten cases. This structure matches the vorticity form of the Navier–Stokes equation. After fixing the structure, we estimated its coefficients for every condition. The advection coefficients remained stable across the series, with mean values of -0.967 and -0.982. By contrast, the two diffusion coefficients varied systematically with the physical setting (Fig. 4a). We therefore treated $c_2$ and $c_3$ as separate targets for dimensionless learning. Dimensionless groups associated with the Reynolds number were removed from the corpus before π-former was retrained. For each diffusion coefficient, the procedure identified:

$$\Pi = \frac{\rho U L}{\mu} \quad (5)$$

which corresponds to the Reynolds number. Power-law fitting produced:

$$c_2 = 0.983\Pi^{-0.993}, \qquad c_3 = 1.051\Pi^{-1.011}. \quad (6)$$

Each closure attained a test $R^2$ of 0.999. Substituting these relations into the discovered structure gave:

$$\omega_t = -0.967u\omega_x - 0.982v\omega_y + 0.983\Pi^{-0.993}\omega_{xx} + 1.051\Pi^{-1.011}\omega_{yy}.$$

The exponents close to -1 recover the inverse-Reynolds scaling of viscous diffusion. Under the adopted sign convention, the advection coefficients also approach their expected values. The ten condition-specific coefficient sets are thereby represented by one equation whose diffusion terms vary through an explicit physical dimensionless number.

We then considered 78 simulations of an advective Fisher–KPP equation[26,27] spanning a two-dimensional space of reaction and transport conditions (Fig. 4b). The concentration fields changed markedly across this space. The genetic search selected the structure:

$$c_t = c_c c + c_{c^2}c^2 + c_{c_x}c_x + c_{c_{xx}}c_{xx} \quad (7)$$

in 33 of the 78 simulations. After fixing this structure, coefficient estimation revealed distinct patterns across the cases. The advection coefficient was concentrated around a mean of -0.999 (Fig. 4b). The linear and quadratic reaction coefficients varied in magnitude while retaining opposite signs. The diffusion coefficient occupied a separate range at lower magnitudes (Fig. 4b). Coefficient-wise dimensionless learning resolved these dependencies. Dimensionless learning identified the same group for the two reaction coefficients, $\Pi_1^* = \frac{U}{Lk}$, whereas the diffusion coefficient optimized $\Pi_2^* = \frac{UL}{D}$. The fitted closures yielded the unified equation:

$$c_t = -0.999c_x + 1.016(\Pi_1^*)^{-0.977}c - 1.024(\Pi_1^*)^{-0.968}c^2 + 0.957(\Pi_2^*)^{-0.992}c_{xx}. \quad (8)$$

The diffusion closure attained $R^2$=0.995, and all three variable coefficients followed exponents close to -1. $\Pi_1^*$ corresponds to the reciprocal of the Damköhler number $\mathrm{Da} = kL/U$. $\Pi_2^*$ corresponds to the Péclet number Pe, the ratio of advection to diffusion. Thus $(\Pi_1^*)^{-1} = \mathrm{Da}$ sets the strength of the logistic source, and $(\Pi_2^*)^{-1} = \mathrm{Pe}^{-1}$ scales the diffusion term. Each variable coefficient is attached to the mechanism that controls it. Even at a noise level of 10%, the workflow recovered the correct equation structure and identified the dimensionless groups governing its variable coefficients, yielding a closed physical equation (Fig. S2).

These examples establish the role of dimensionless learning in equation discovery. Structure discovery identifies the terms that recur across conditions. The variation of their coefficients then reveals the missing closure. Expressing this variation through discovered dimensionless numbers converts a collection of case-specific estimates into a single governing equation that changes predictably with the physical setting. Dimensionless numbers thereby become part of the discovered law itself.

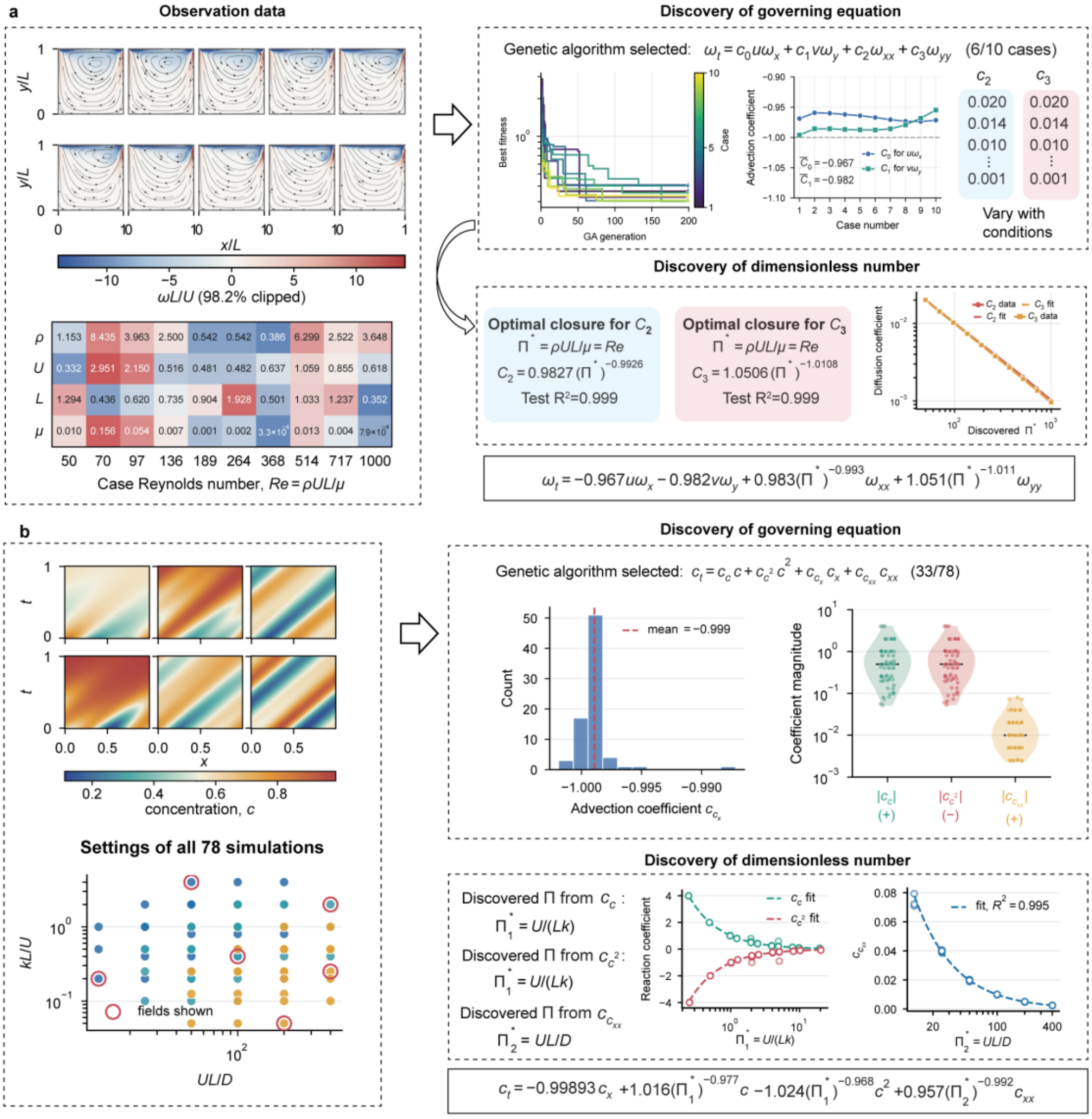

**Fig. 4. Discovered dimensionless closure of condition-dependent equations. (a),** Lid-driven cavity flow across ten Reynolds-number conditions. Vorticity fields and simulation settings (left), the common equation optimized by the genetic algorithm and its fitted coefficients (upper right), and dimensionless closures for the diffusion coefficients (lower right). **(b),** Reaction-advection-diffusion dynamics across 78 parameter settings. Representative concentration fields and simulation conditions (left), the optimized equation and coefficient distributions (upper right), and dimensionless closures for the reaction and diffusion coefficients (lower right).

**Discovering physical closure from real-world evaporation-driven wetting experiments**

We next applied the framework to real-world experimental data from evaporation-driven wetting of a thin porous medium. After a finite-volume droplet enters the pores, capillary transport drives radial spreading while evaporation removes liquid from the exposed surfaces. The wet-patch radius therefore increases to a maximum $R^*$ at time $t^*$ and subsequently contracts until complete drying. This process is important to support radius-based volume inference in bloodstain analysis and pesticide dosing[28]. Existing scaling relations cover restricted regimes, and some contain dimensional inconsistencies. A common description across liquids and initial droplet volumes therefore remains an open problem.

We analyzed the experiments reported by Lee et al.[29] on suspended Whatman 41 filter paper with a thickness of $h$=215 $\mu$m, allowing evaporation from its upper and lower surfaces. Time zero marked complete absorption of the droplet, when the wet-patch radius was $R_0$. The subsequent radius $R(t)$ was recorded until drying. The dataset contained 24 cases involving hexane, heptane, octane, nonane and dodecane (Fig. 5a). Evaporation was represented by $j_0$, consistent with $\mathrm{d}V/\mathrm{d}t=-j_0 R$. Capillary transport involved the viscosity $\mu$ and surface tension $\gamma$. The substrate was described by its thickness $h$ and microstructural length $l=k_s\kappa$, which was approximately 50 nm for the filter paper. From these variables, the $\pi$-former generated three basic groups:

$$\Pi_1 = \frac{j_0\mu}{R_0\gamma}, \qquad \Pi_2 = \frac{R_0}{h}, \qquad \Pi_3 = \frac{R_0^2}{h\ell}. \tag{9}$$

Here, $\Pi_1$ measures evaporation relative to capillary transport. $\Pi_2$ describes the aspect ratio of the initial wet patch. $\Pi_3$ compares the macroscopic patch area with the microstructural transport scale. Data-driven optimization then produced

$$\Pi_R = (\Pi_1\Pi_3)^{1/2} = \left(\frac{j_0\mu R_0}{\gamma h\ell}\right)^{1/2} \tag{10}$$

for the maximum spreading ratio $R^*/R_0$ and

$$\Pi_t = \Pi_1^{1/2} = \left(\frac{j_0\mu}{R_0\gamma}\right)^{1/2} \tag{11}$$

for the characteristic time $t^*/t_{ref}$, $t_{ref}$=1s (Fig. 5b). The power-law relation between $R^*/R_0$ and $\Pi_R$ had an exponent of -0.301 and $R^2$=0.979. The relation between characteristic time $t^*/t_{ref}$ and $\Pi_t$ had an exponent of -2.136 and $R^2$=0.994 (Fig. 5c). These dimensionless numbers accounted for most of the variation arising from liquid properties and geometric scales.

We next defined $x=t/t^*$ and $y=R/R^*$ as new coordinates, and used the graph-based equation discovery method (GraphED)[3] to discover a common expression for the normalized trajectories (Fig. 5d). More details about the method can be found in Supplementary Information S.3. The equation discovery process identified a previously unexplored form:

$$y = C_4(1 + C_6x)[(1 - x)^2 - C_3x^{C_8}]^{1/2}, \tag{12}$$

where $C_3$, $C_4$, $C_6$ and $C_8$ are free parameters (Fig. 5d). This expression reproduced the measured trajectories, while its behaviour near the maximum remained open to physical refinement. The definitions of $R^*$ and $t^*$ impose $y(1)=1$ and $y'(1)=0$. Enforcing these conditions reduced the equation

to

$$y = \left(1 + \frac{p}{2} - \frac{px}{2}\right)[x^p + c(x-1)^2]^{1/2}, \tag{13}$$

with two free parameters, $c$ and $p$. The resulting equation represents the increase towards the maximum radius and the subsequent evaporation-driven contraction through one continuous relation. The discovered equation achieved $R^2$=0.994 and an RMSE of 0.012 on the training cases (Fig. 5e). The held-out set comprised all heptane cases and one additional nonane droplet-size condition. The corresponding predictions retained $R^2$=0.994 and an RMSE of 0.010. The equation therefore preserved its accuracy across an unseen liquid and a held-out initial size. It also satisfies the defining properties of the turning point, where the normalized radius reaches unity and its rate of change vanishes.

A single global curve captured the initial spreading stage and the longer contraction stage for different liquids with $R^2$=0.767 (Fig. 5f). The empirical relation proposed previously gave $R^2$=0.683 on the same normalized trajectories (Fig. S3)[29]. The framework thus identified dimensionless numbers as physical coordinates and discovered a compact equation from real experimental data, which outperforms the referenced empirical model. More analysis and comparisons are provided in Supplementary Information S.4.

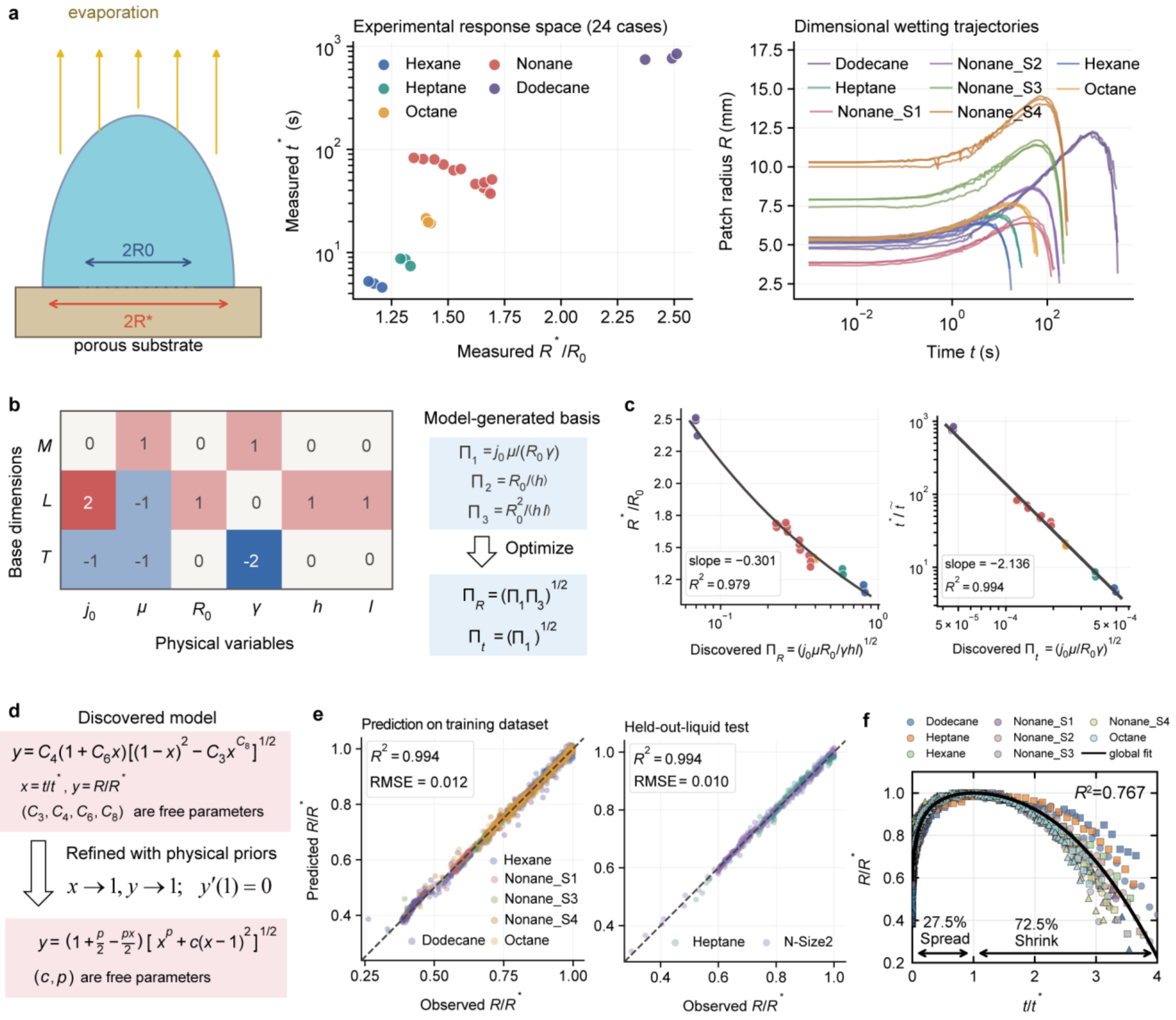

**Fig. 5. Discovery of dimensionless wetting dynamics from experimental data. (a),** Experimental system, response space and wet-patch trajectories. **(b),** Dimensional encoding, generated basis dimensionless groups and optimized dimensionless numbers. **(c),** Scaling relations for the maximum spreading ratio and characteristic time. **(d),** Discovered trajectory equation and its refinement using physical constraints. **(e),** Predictions for the training and held-out cases. **(f),** Normalized trajectories and the global fit.

## Discussion

This work addresses a central ambiguity in dimensionless analysis. Dimensional homogeneity defines the admissible space, while the physically useful dimensionless groups remain underdetermined. We introduced a framework that utilizes a generative model called π-former to learn how scientists have constructed dimensionless groups. On this basis, the Buckingham constraint is enforced during generation and observations are used to optimize the combinations for a target phenomenon. The selected dimensionless numbers then provide explicit closures for coefficients in discovered equations. Across simulated and real-world experiments, the framework recovered established dimensionless numbers and produced closed relations for condition-dependent dynamics. Its main contribution is an alternative route from accumulated scientific knowledge and data to testable dimensionless hypotheses and physical equations.

The results distinguish the physical relevance of a dimensionless group from its predictive contribution within a particular dataset. In Rayleigh–Bénard convection, π-former proposed the Prandtl number despite its nearly constant value in the available observations. Its exclusion from the fitted relation therefore reflects the sampled conditions, while its presence among the generated hypotheses preserves a potentially relevant physical dependence. Such hypotheses may help identify which parameters to vary in subsequent experiments and where an inferred relation requires further testing. The value of the learned prior thus extends beyond search efficiency to retaining physical possibilities that the current data have limited capacity to distinguish. Integrating dimensionless learning with equation discovery gives condition-dependent coefficients an explicit physical interpretation. Expressing these coefficients through discovered dimensionless groups yields a unified equation across operating conditions, connecting hypothesis generation with empirical testing and physical closure. This integration also illustrates a broader role for scientific literature in machine learning. Published dimensionless groups encode accumulated experience in physical abstraction, which can be learned and transferred to new discovery tasks. Scientific knowledge thereby guides the proposed hypotheses, while observations determine which enter the final equation.

Several limitations define the present scope. The corpus was drawn from a single compilation and therefore reflects its disciplinary coverage and historical selection. Recovery also declined as the number of candidate variables increased. The discrete exponent vocabulary favours familiar rational forms, while unusual exponents and non-monomial coordinates require a broader representation. Subset evaluation becomes increasingly expensive for large variable sets. Future work should expand the corpus through literature extraction. Active experimental design could then select conditions that best distinguish candidate groups. Joint inference of equation structure and dimensionless closure may further extend the framework to systems in which these two components evolve together.

## Materials and Methods

The proposed framework separates dimensionless learning into four stages (Fig. 1). Literature-derived dimensionless numbers are first converted into set-valued training sets. The π-former then learns a conditional distribution over the exponent assigned to each variable. At inference, this learned distribution guides a discrete search under the exact Buckingham constraint. The resulting basic groups span the admissible dimensionless space and retain simple algebraic forms. Measurements from the target problem finally select one or more composite dimensionless numbers, denoted by $\Pi^*$. A separate procedure uses the discovered dimensionless numbers to close coefficients in equations discovered across operating conditions.

### Encoding literature-derived dimensionless knowledge

We assembled a corpus of 1,288 dimensionless numbers reported across scientific and engineering fields. Each record contains a variable set

$$\mathrm{X}^{(g)} = \{x_1^{(g)}, \ldots, x_{m_g}^{(g)}\} \tag{14}$$

and its dimensional encoding. The exponent vector in the published expression provides the token-level labels. Scalar constants were removed because they leave dimensional homogeneity unchanged. A dimensionless group $g$ was written as

$$\Pi^{(g)} = \prod_{i=1}^{m_g} \left(x_i^{(g)}\right)^{\alpha_i^{(g)}}. \tag{15}$$

The dimensions of each variable were expressed in the seven-dimensional SI basis

$$\mathbf{b} = (M, L, T, I, \Theta, N, J). \tag{16}$$

The corresponding vector $\mathbf{d}_i^{(g)} \in \mathrm{Q}^7$ records the power of each base dimension. Placing these vectors as columns gives the dimension matrix

$$D^{(g)} = \left[\mathbf{d}_1^{(g)} \quad \cdots \quad \mathbf{d}_{m_g}^{(g)}\right] \in \mathrm{Q}^{7\times m_g}. \tag{17}$$

Dimensional homogeneity requires

$$D^{(g)}\boldsymbol{\alpha}^{(g)} = \mathbf{0}. \tag{18}$$

Each entry of $D^{(g)}\boldsymbol{\alpha}^{(g)}$ is the total exponent of one base dimension in the resulting product. The zero vector therefore guarantees invariance under changes of measurement units.

Exponent vectors that differ by a common scale describe powers of the same coordinate. We mapped each literature expression to a canonical representative before training. Rational exponents were converted to their smallest integer ratio, after which a fixed sign convention removed reciprocal duplicates. The stored labels were then rescaled to the reporting convention used in the corpus. This treatment makes equivalent expressions share one supervised target while preserving fractional exponents in the final formula. Each variable was represented by a token containing its dimensional vector. A training example therefore took the form

$$\mathrm{S}^{(g)} = \left\{\left(x_i^{(g)}, \mathbf{d}_i^{(g)}\right)\right\}_{i=1}^{m_g}, \tag{19}$$

with the canonical exponents as token-level labels. The records have different cardinalities, so padded elements were excluded through an attention mask. Dataset partitions were defined at the level of complete dimensionless groups. This split preserves the held-out reconstruction task used in Fig. 2.

**Learning exponent distributions with the $\pi$-former**

The π-former maps an unordered variable set to a probability distribution for each exponent. Its backbone is a set transformer composed of stacked self-attention blocks. Self-attention allows the prediction for one variable to depend on the other variables present in the same group. The representation remains permutation equivariant, so changing the input order changes only the corresponding output order. For a set S, the network produces token-wise probabilities

$$p_\theta(\alpha_i = a \mid \mathrm{S}), \qquad a \in \mathrm{A}, \tag{20}$$

where A is the finite exponent vocabulary obtained from the canonicalized corpus. The output represents an empirical prior over exponent patterns observed in scientific usage. Dimensional validity is imposed during decoding, which leaves the network free to learn variable associations and preferred exponent magnitudes from the literature. The parameters $\theta$ were fitted using masked token-level cross-entropy:

$$\mathrm{L}_\pi(\theta) = -\frac{1}{G}\sum_{g=1}^{G}\frac{1}{m_g}\sum_{i=1}^{m_g}\log p_\theta\left(\alpha_i^{(g)} \mid \mathrm{S}^{(g)}\right), \tag{21}$$

where $G$ is the number of training groups. Each token contributes one exponent label, while self-attention conditions that prediction on the full variable set. The model can therefore learn recurring construction patterns. Examples include the tendency of transport coefficients to enter denominators and the use of simple rational powers for characteristic scales. The learned distribution is used as a prior score during discovery. It carries information accumulated across published problems, while the subsequent physical constraint retains only dimensionally admissible assignments. This separation prevents predictive uncertainty from weakening dimensional homogeneity. The π-former settings used in this study are detailed below.:

The encoder contained three set attention blocks with a hidden dimension of 128. Each block used four attention heads and a $128 \rightarrow 512 \rightarrow 128$feed-forward network with ReLU activation. Dropout was set to 0.1 within the attention blocks. The classification head mapped each contextualized variable through $128 \rightarrow 128 \rightarrow 39$, with ReLU activation after the first layer. A separate softmax distribution over 39 exponent classes was produced for each variable. Padding masks were applied during attention, and padded labels were assigned -100. The model contained 617,383 trainable parameters. Training used AdamW for 3,000 epochs with a learning rate of $10^{-3}$and weight decay of $10^{-4}$. We use full-batch training for the 1,288 samples. The learning rate remained fixed, and the checkpoint with the lowest mean training loss was retained. Training used CUDA when available. Cross-validation partitions used a seed of 0, while the main optimization runs retained stochastic initialization.

**Constrained generation of basic dimensionless groups**

Consider a target problem with variables $\mathrm{X}=\{x_1,\ldots,x_n\}$and dimension matrix $D\in\mathrm{Q}^{7\times n}$. Buckingham's theorem gives $K = n\text{-rank}(D)$ independent dimensionless groups. The theorem specifies the dimension of this space, while leaving its basis open. We use the learned prior to rank simple bases inside this null space. The decoder enumerates informative subsets $S\subseteq\{1,\ldots,n\}$. For each subset, $D_S$ denotes the submatrix formed by its variables. Subsets with positive nullity enter constrained decoding.

The π-former outputs $p_\theta(\alpha_i|S)$, after which the exponent vector is obtained from

$$\widehat{\boldsymbol{\alpha}}_S = \underset{\boldsymbol{\alpha}_S\in\mathrm{A}^{|S|}}{\mathrm{argmax}}\,\mathrm{score}(S, \boldsymbol{\alpha}_S) \tag{22}$$

subject to

$$D_S\boldsymbol{\alpha}_S=\mathbf{0},\qquad \boldsymbol{\alpha}_S\neq\mathbf{0}. \tag{23}$$

The candidate score is

$$\mathrm{score}(S,\boldsymbol{\alpha}_S)=\sum_{i\in S}\log p_\theta(\alpha_i\mid S)-\lambda_S|S|-\lambda_\alpha(\|\boldsymbol{\alpha}_S\|_1+0.1\,\|\boldsymbol{\alpha}_S\|_0). \tag{24}$$

The first term measures agreement with the learned construction prior. The subset penalty favours groups involving fewer variables. The final penalty favours small exponents and sparse support. Together, these terms prefer formulas that resemble established dimensionless numbers while retaining a compact algebraic form.

Decoding is performed in a discrete candidate pool formed from the highest-probability exponent labels at each token. Every candidate is checked against $D_S\boldsymbol{\alpha}_S=\mathbf{0}$ using exact rational arithmetic. The pool is expanded when it yields too few admissible vectors. Surviving vectors are first canonicalized. Variables outside $S$ are then zero-padded to embed each vector in $\mathrm{Q}^n$. Duplicate vectors across subsets are merged. Each vector defines

$$\Pi(\boldsymbol{\alpha})=\prod_{i=1}^{n}x_i^{\alpha_i}. \tag{25}$$

Candidates are sorted by their scores. A candidate is retained when its exponent vector increases the rank of the selected set. The procedure ends after $K$ linearly independent vectors have been obtained:

$$\mathrm{rank}[\boldsymbol{\alpha}_1\ \cdots\ \boldsymbol{\alpha}_K]=K. \tag{26}$$

The resulting set $\{\Pi_1,\dots,\Pi_K\}$ forms a basis of the Buckingham null space. The basis is dimensionally complete, yet its specific form reflects patterns learned from the literature. These groups are treated as interpretable building blocks for the data-driven stage.

**Data-driven optimization of dimensionless groups**

The π-former identifies plausible coordinates, while measurements determine which dimensionless number describes the target response. For observations $\{(\boldsymbol{x}^{(q)},y^{(q)})\}_{q=1}^{N}$, we first evaluate every basic group $\Pi_k$ on each sample. A composite coordinate is written as

$$\Pi^*(\boldsymbol{\beta})=\prod_{k=1}^{K}\Pi_k^{\beta_k}, \tag{27}$$

or equivalently,

$$\log\Pi^*(\boldsymbol{\beta})=\sum_{k=1}^{K}\beta_k\log\Pi_k. \tag{28}$$

The coefficients $\boldsymbol{\beta}$ are searched over a bounded set of simple rational values. This representation permits inversion and multiplication of the basic groups while keeping the final expression readable. Here, the default set is [-3, -2, -3/2, -1, -1/2, 0, 1/2, 1, 3/2, 2, 3]. For each $\boldsymbol{\beta}$, a response model $f_m$ is fitted between $\Pi^*(\boldsymbol{\beta})$ and $y$. The index $m$ denotes model complexity. We begin with a linear relation and increase the polynomial degree only when the validation score remains below the prescribed threshold. Selection minimizes

$$\mathrm{J}_{\mathrm{data}}(\boldsymbol{\beta}, m) = 1 - R^2 + \lambda_\beta(\| \boldsymbol{\beta} \|_1 + 0.1 \| \boldsymbol{\beta} \|_0) + \lambda_m \mathrm{C}(f_m), \tag{29}$$

where $\mathrm{C}(f_m)$ increases with model order. The selected coordinate and response model are

$$(\boldsymbol{\beta}^*, m^*) = \underset{\boldsymbol{\beta}, m}{\operatorname{argmin}} \mathrm{J}_{\mathrm{data}}(\boldsymbol{\beta}, m), \tag{30}$$

with

$$\hat{y} = f_{m^*}\left(\Pi^*(\boldsymbol{\beta}^*)\right). \tag{31}$$

This objective balances predictive adequacy against the complexity of the discovered dimensionless number. It also penalizes unnecessary flexibility in the response model. Validation performance is evaluated on held-out samples.

Some responses require more than one dimensionless coordinate. In this setting, the scalar search is performed first. If its validation score remains below the target level, a second coordinate is introduced:

$$\Pi_r^* = \prod_{k=1}^{K} \Pi_k^{\beta_{rk}}, \qquad \hat{y} = f_m(\Pi_1^*, \ldots, \Pi_R^*). \tag{32}$$

The exponent vectors $\boldsymbol{\beta}_r$ are required to be linearly independent. Coordinates are added sequentially, and each addition must improve held-out performance enough to offset its complexity penalty.

**Closing discovered equations with dimensionless learning**

For equation discovery, the target of dimensionless learning is a coefficient that varies across cases. Suppose a shared operator structure has been identified from $C$ operating conditions:

$$\partial_t u^{(c)} = \sum_{j=1}^{P} a_j^{(c)} \phi_j(u^{(c)}, \nabla u^{(c)}, \ldots), \qquad c = 1, \ldots, C. \tag{33}$$

Here, we utilize partial differential equations as examples. $\phi_j$ denotes a discovered operator and $a_j^{(c)}$ is its fitted coefficient in case $c$. We used genetic structure optimization to identify recurrent terms, followed by regression on each case to estimate the corresponding coefficients. Coefficients with stable values across conditions were pooled into a shared estimate $\bar{a}_j$. Coefficients showing systematic variation were passed to the closure stage. Each case has a vector of physical parameters $\mathbf{x}^{(c)}$ and therefore a value for every basic group. For a varying coefficient $a_j$, the framework searches for

$$\Pi_j^{*(c)}(\boldsymbol{\beta}_j) = \prod_{k=1}^{K} \left(\Pi_k^{(c)}\right)^{\beta_{jk}} \tag{34}$$

and a low-complexity closure function $g_j$. The dimensionless coefficient is then used as the response in the closure search. Case-level splits are used for validation, so the learned closure is assessed on operating conditions excluded from fitting. Substituting the fitted closures into the shared structure gives

$$\partial_t u = \sum_{j\in I} \bar{a}_j \phi_j + \sum_{j\in V} s_j(\mathbf{x}) g_j^* \left(\Pi_j^*(\mathbf{x})\right) \phi_j, \tag{35}$$

where $I$ contains condition-invariant coefficients and $V$ contains coefficients with dimensionless closure. The final expression retains one operator structure across cases. Its coefficients change through explicit physical coordinates, which turns cross-case variation into part of the discovered equation.

**Acknowledgments:**
This work was supported and partially funded by the National Natural Science Foundation of China (grant no. 52288101, 12501744). This work was supported by the High Performance Computing Centers at the Eastern Institute of Technology, Ningbo, and the Ningbo Institute of Digital Twin.

**Author contributions**
H.X., T.W., and D.Z. conceived the idea, designed the study, and analyzed the results. H.X. developed the algorithm, performed the computations, and generated the results and figures. T.W. and D.Z. supervised the project. All of the authors contributed to the writing and editing of the manuscript.

**Competing interests**
Authors declare that they have no competing interests.

**Data availability**
All datasets generated in this work have been deposited in the open-source repository: https://github.com/woshixuhao/Discovery_of_dimensionless_groups.

**Code availability**

All codes generated in this work have been deposited in the open-source repository: https://github.com/woshixuhao/Discovery_of_dimensionless_groups.

# **Supplementary Information** for

# Discovering like a scientist: Literature-informed dimensionless learning for physical closure of data-driven equations

Hao Xu[1], Tianhao Wu[1,*], and Dongxiao Zhang[1,*]

[1] Zhejiang Key Laboratory of Industrial Intelligence and Digital Twin, Eastern Institute of Technology, Ningbo, Zhejiang 315200, China.

[*] Corresponding authors.

Email address: twu@eitech.edu.cn (T. Wu); dzhang@eitech.edu.cn (D. Zhang).

## Supplementary text

### S.1 The introduction and settings for benchmark cases utilized in this study

The seven cases below follow the IT-$\pi$ suite. Sample size $N$ is the number of rows in $X$. Unless noted, the observed output is already dimensionless; the learning task is to recover the input-side dimensionless groups. All comparisons use an 80/20 split with seed 42.

**Keyhole** Laser melting of a metal produces a vapour cavity whose depth $e$ is predicted in the dimensionless form $e^* = e/r_0$. The seven inputs are the absorbed laser power $\eta P$ (W), scan speed $V_s$ (m·s$^{-1}$), density $\rho$ (kg·m$^{-3}$), specific heat $c_p$ (J·kg$^{-1}$·K$^{-1}$), melting-to-ambient temperature difference $T_l - T_0$ (K), spot radius $r_0$ (m) and thermal diffusivity $\alpha$ (m$^2$·s$^{-1}$).

**Dixit** Smooth-wall turbulent boundary layers under zero, adverse and favourable pressure gradients. The six inputs are free-stream speed $U_\infty$ (m·s$^{-1}$), kinematic viscosity $\nu$ (m$^2$·s$^{-1}$), a pressure-gradient parameter $M$(m$^3$·s$^{-2}$), boundary-layer thickness $\delta$ (m), displacement thickness $\delta^*$ (m) and momentum thickness $\theta$ (m). The target is the friction velocity $u_\tau$ (m·s$^{-1}$), the only dimensional output in the suite.

**Rayleigh-Bénard** A fluid layer of height $h$ is heated from below. The seven inputs are $h$ (m), temperature difference $\Delta T$ (K), thermal conductivity $\lambda$ (W·m$^{-1}$·K$^{-1}$), gravity $g$ (m·s$^{-2}$), thermal expansion coefficient $\alpha$ (K$^{-1}$), kinematic viscosity $\nu$ (m$^2$·s$^{-1}$) and thermal diffusivity $\kappa$ (m$^2$·s$^{-1}$). The target is the Nusselt number Nu. The classical coordinates are

$$\mathrm{Ra} = \frac{g\alpha\Delta T h^3}{\nu\kappa}, \qquad \mathrm{Pr} = \frac{\nu}{\kappa}. \tag{S.1}$$

**MHD generator** A conducting fluid is driven along a channel by a streamwise pressure gradient and opposed by a Lorentz force. The six inputs are channel half-width $l$ (m; set to 1 in the file), dynamic viscosity $\mu$ (Pa·s), density $\rho$ (kg·m$^{-3}$), electrical conductivity $\eta$ (S·m$^{-1}$), pressure gradient $\mathrm{d}p/\mathrm{d}x$ (Pa·m$^{-1}$) and magnetic field $B$ (T). The raw mean speed $U_{\mathrm{avg}}$is converted to $\mathrm{Re}_u = \rho U_{\mathrm{avg}} l/\mu$, which is the regression target.

**Blasius** Steady laminar flow past a semi-infinite flat plate at zero incidence. The four inputs are free-stream speed $U$ (m·s$^{-1}$; fixed at 0.01 in the synthesis), kinematic viscosity $\nu$ (m$^2$·s$^{-1}$; fixed at $10^{-6}$), streamwise station $x$ (m) and wall-normal station $y$ (m). The target is the velocity ratio $u/U$, which depends on the similarity variable

$$\eta = y\left(\frac{U}{\nu x}\right)^{1/2}. \tag{S.2}$$

**Colebrook** Turbulent friction in a rough circular pipe. The five inputs are mean speed $u$ (m·s$^{-1}$), density $\rho$ (kg·m$^{-3}$; fixed at 200), diameter $D$ (m), absolute roughness $k$ (m) and dynamic viscosity $\mu$ (Pa·s; fixed at 0.001). The target is the Darcy friction factor $f$ generated from the Colebrook-White relation, which depends on

$$\mathrm{Re} = \frac{\rho u D}{\mu}, \quad \frac{k}{D}. \tag{S.3}$$

**Rayleigh (Stokes' first problem)** An infinite plate is started impulsively at speed $U$ in a quiescent viscous fluid. The four inputs are $U$ (m·s$^{-1}$), wall-normal distance $y$ (m), time $t$ (s) and kinematic viscosity $\nu$ (m$^2$·s$^{-1}$). The target follows

$$\frac{u}{U} = 1 - \mathrm{erf}\left(\frac{y}{2\sqrt{\nu t}}\right), \tag{S.4}$$

and depends on $y/\sqrt{\nu t}$.

## S.2 Sensitivity analysis of the candidate exponent set in data-driven optimization

We examined how the resolution of the candidate exponent set in data-driven optimization affects the computational cost and final dimensionless groups. All five settings covered the interval [−3,3], while their cardinality and denominator resolution varied. The tested sets were

$$\mathcal{B}_{\text{simplified}} = \{k \mid k \in \mathbb{Z}, -3 \le k \le 3\}, \quad |\mathcal{B}_{\text{simplified}}| = 7,$$

$$\mathcal{B}_{\text{original}} = \mathcal{B}_{\text{simplified}} \cup \left\{\pm\frac{1}{2}, \pm\frac{3}{2}\right\}, |\mathcal{B}_{\text{original}}| = 11,$$

$$\mathcal{B}_{\text{enhanced-1}} = \mathcal{B}_{\text{original}} \cup \left\{\frac{k}{3} \mid k = -9, \dots, 9\right\}, \quad |\mathcal{B}_{\text{enhanced-1}}| = 23,$$

$$\mathcal{B}_{\text{enhanced-2}} = \mathcal{B}_{\text{enhanced-1}} \cup \left\{\frac{k}{5} \mid k = -15, \dots, 15\right\}, \quad |\mathcal{B}_{\text{enhanced-2}}| = 47,$$

$$\mathcal{B}_{\text{enhanced-3}} = \left\{\frac{k}{10} \mid k = -30, \dots, 30\right\}, |\mathcal{B}_{\text{enhanced-3}}| = 61.$$

The computational time increased with the number of admissible exponents because each additional value enlarged the space of candidate combinations (Supplementary Fig. S1). The simplified and original sets completed most cases within a few seconds. The finer sets required several to tens of seconds. Even the denominator-10 grid completed all seven problems within approximately 70 s. It therefore provides a resolution of 0.1 while retaining a minute-scale search cost. Expanding the candidate set preserved the selected dimensionless groups in most of the seven problems. This result indicates that integer and half-integer exponents capture the dominant dimensional groups in many established cases. Additional nonlinear dependence can then be represented by the fitted scaling law. A denser exponent set remains useful for open problems with less established structure. In the Keyhole case, expansion to the Enhanced-2 set produced a dimensionless relation with a test $R^2$ of 0.981. The search can therefore begin with the original set and progressively introduce finer rational exponents when the predictive criterion remains unmet. This strategy balances computational cost with the capacity to resolve more complex dimensionless coordinates.

### S.3 The settings of equation discovery methods.

For the lid-driven cavity flow, the left-hand side was fixed as $\omega_t$. Basic genes for right-hand-side terms were $\{u, v, \omega, \omega_x, \omega_y, \omega_{xx}, \omega_{yy}\}$. The genetic algorithm used 400 individuals for 200 generations and allowed up to five terms per equation. Candidate coefficients were estimated by least squares. Fitness was defined as

$$F = MSE + \epsilon L, \tag{S.5}$$

where $L$ denotes the total equation length and $\epsilon$=0.05. The best half of each population was retained, while the remaining individuals were resampled. Crossover exchanged one term between paired genomes in every generation. The probabilities of term addition, term deletion and factor mutation were 0.4, 0.5 and 0.4, respectively. The elite genome was preserved. Derivatives were evaluated using a seven-point finite-difference scheme on a 40×40×40 grid after removing eight spatial points and four temporal points from each boundary. The random seed was 525.

The advective Fisher–KPP equation case used the basic genes $\{c, c_x, c_{xx}, c_{xxx}\}$. Each candidate contained at most five terms. The algorithm compared $c_t$ and $c_{tt}$ as possible left-hand-side terms. Coefficients were obtained from the smallest singular vector and normalized by the selected left-hand-side coefficient. Fitness was the same as (Eq. S.5), with $\epsilon=2\times10^{-5}$. Surrogate models are constructed by artificial neural networks and the derivatives are calculated by the automatic differentiation. The optimization used 400 individuals for 200 generations. The addition and point-mutation rates were 0.4 and 0.3, respectively; term deletion also used a rate of 0.3. Ten boundary points were removed before residual evaluation. Selection retained the best half of the population and filled the remainder with randomly sampled genomes. Crossover exchanged one term between every paired genome, and the leading genome was preserved. The random seed was 525.

The normalized wetting trajectories were modelled using GraphED, which represents mathematical expressions as directed graphs. The operator set included addition, multiplication, exponential and logarithmic functions. Edge attributes marked as free coefficients were optimized separately for each trajectory using BFGS. Twenty initial coefficient sets were screened before each optimization. The search used 300 graphs for 150 generations. Each graph contained at most 30 edges, seven variable nodes and four free coefficients. Fitness was the mean MSE across the training trajectories. Offspring were merged with the parent population, deduplicated and ranked; the best half was retained and the population was completed with newly generated graphs. The mutation rates were 0.3 for nodes, 0.5 for subgraphs, 0.2 for subgraph deletion and 0.5 for edge attributes. The population was reinitialized after five generations of unchanged best fitness, with the leading graph retained. Computation used all available CPU cores except one. Python and NumPy seeds were set to 1101 and 525, respectively. Training used the dodecane, hexane, nonane and octane trajectories. Heptane and the second nonane-size condition formed the held-out set.

## S.4 Analysis of the discovered closed physical equation for evaporation-driven wetting

Lee et al.[29] developed a mechanistic model for evaporation-driven radial wicking in thin porous media. The model combines Darcy transport with a saturation-dependent permeability $k = k_s S^3$, a capillary-pressure gradient $\mathrm{d}P/\mathrm{d}r = \kappa\gamma/r$, and the measured evaporation law $\mathrm{d}V/\mathrm{d}t = -j_0 R$. Defining $l = k_s\kappa$ and the spreading timescale:

$$t_s = \frac{\mu\phi R_0^2}{6l\gamma}, \tag{S.6}$$

the authors introduced the evaporation number

$$E = \frac{j_0\mu R_0}{4\pi\gamma h l}. \tag{S.7}$$

An asymptotic solution for the late-stage dynamics gave

$$\tilde{R} \simeq \left(1 - \frac{4}{7}E\tilde{t}^{7/6}\right)^{1/2} \tilde{t}^{1/6}, \tag{S.8}$$

from which the maximum radius and its occurrence time follow as

$$\frac{R^*}{R_0} = \left(\frac{7}{9}\right)^{1/2}\left(\frac{7}{18}\right)^{1/7} E^{-1/7}, \qquad \frac{t^*}{t_s} = \left(\frac{7}{18}\right)^{6/7} E^{-6/7}. \tag{S.9}$$

With $x = t/t^*$ and $y = R/R^*$, this analysis produced the parameter-free trajectory

$$y_{\mathrm{Lee}}(x) = \frac{3}{\sqrt{7}}\left(1 - \frac{2}{9}x^{7/6}\right)^{1/2} x^{1/6}. \tag{S.10}$$

Our framework recovered the same governing balance directly from the candidate variables. The generated basic groups were

$$\Pi_1 = \frac{j_0\mu}{R_0\gamma}, \qquad \Pi_2 = \frac{R_0}{h}, \qquad \Pi_3 = \frac{R_0^2}{hl}. \tag{S.11}$$

Data-driven optimization selected

$$\Pi_R = (\Pi_1\Pi_3)^{1/2} = \left(\frac{j_0\mu R_0}{\gamma h l}\right)^{1/2} = (4\pi E)^{1/2} \tag{S.12}$$

for the maximum spreading ratio. The fitted relation $R^*/R_0 \propto \Pi_R^{-0.301}$ corresponds to $R^*/R_0 \propto E^{-0.1505}$, close to the theoretical exponent $E^{-1/7}$. For the characteristic time, the framework independently selected

$$\Pi_t = \Pi_1^{1/2} = \left(\frac{j_0\mu}{R_0\gamma}\right)^{1/2}, \tag{S.13}$$

which gave $t^* \propto \Pi_t^{-2.136}$ with $R^2 = 0.994$. The previous model applies one manually derived evaporation number after normalization by $t_s$. Our procedure selects a response-specific coordinate from literature-informed hypotheses and experimental observations.

Graph-based equation discovery further produced a unified trajectory. Applying the physical conditions $y(1) = 1$ and $y'(1) = 0$ yielded

$$y = \left(1 + \frac{p}{2} - \frac{px}{2}\right)[x^p + c(x-1)^2]^{1/2}, \tag{S.14}$$

where $c$ and $p$ are fitted parameters. On the common global dataset excluding hexadecane, the Lee *et al.* relation attained $R^2 = 0.683$, whereas the discovered equation attained $R^2 = 0.767$. The reference relation provides a parameter-free prediction based on prescribed transport assumptions. The discovered equation preserves the peak conditions while adapting the trajectory shape to the measured dynamics. The discovered equation is able to assign liquid-specific values of $c$ and $p$, accurately reproducing the trajectory of each liquid. A single global fit across all liquids also outperformed the existing empirical relation, demonstrating the value of integrating dimensionless discovery with equation discovery for real experimental systems.

## Supplementary tables and figures

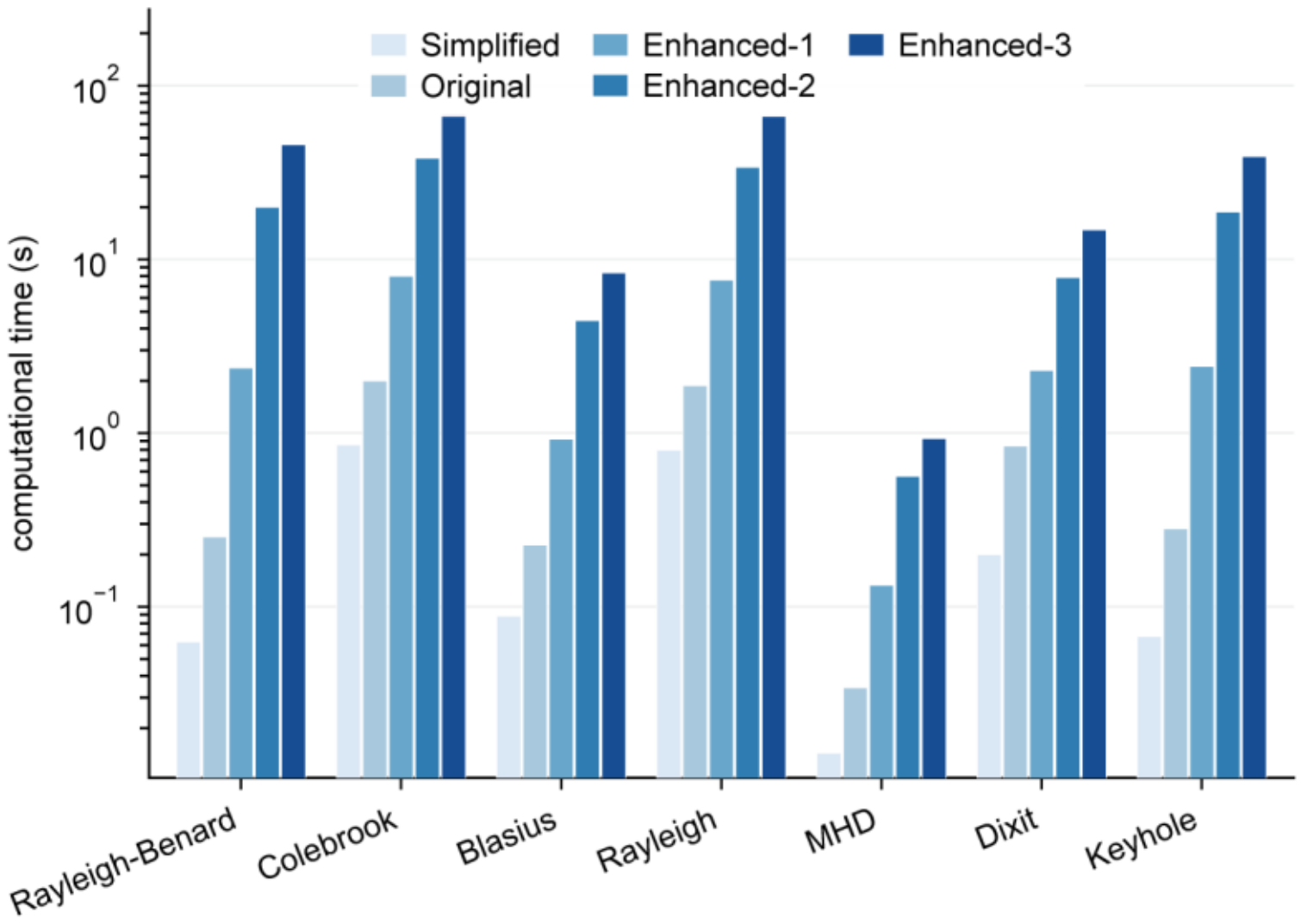

**Fig. S1. Computational time across benchmark cases.** Wall-clock time for the simplified, original and three enhanced search settings for the range of exponents $\beta$. The vertical axis uses a logarithmic scale.

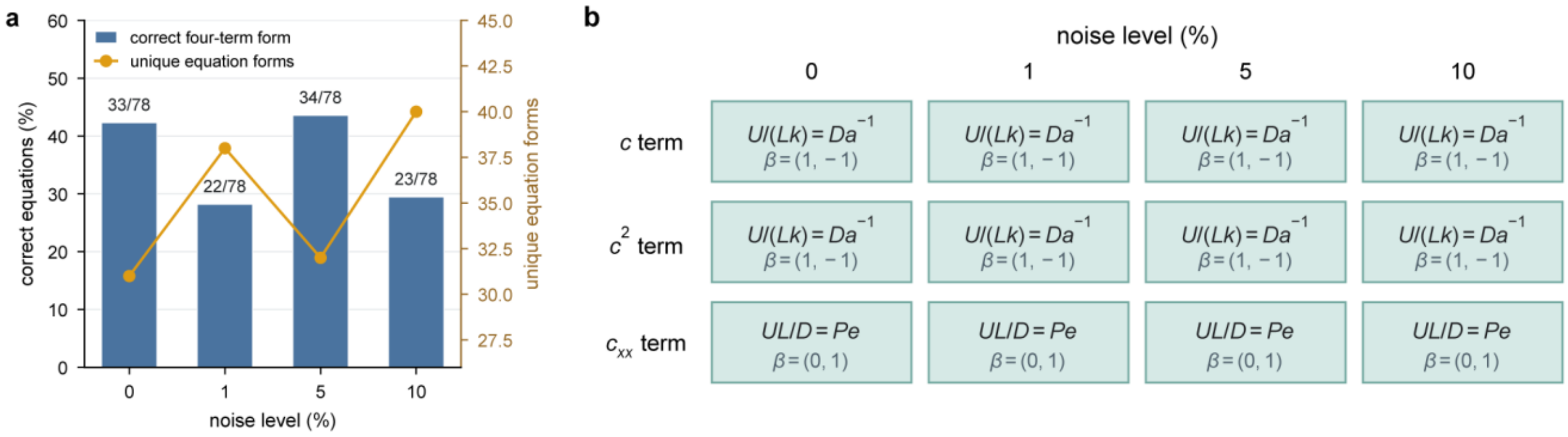

**Fig. S2. Noise sensitivity of equation and dimensionless discovery. (a)**, Recovery rate of the correct four-term advective Fisher-KPP equation across 78 cases (blue bars) and number of unique equation forms (orange line) at noise levels of 0, 1, 5 and 10%. **(b)**, Dimensionless numbers optimized for the coefficients of the discovered structure under different noise. Here, the noise is Gaussian noise.

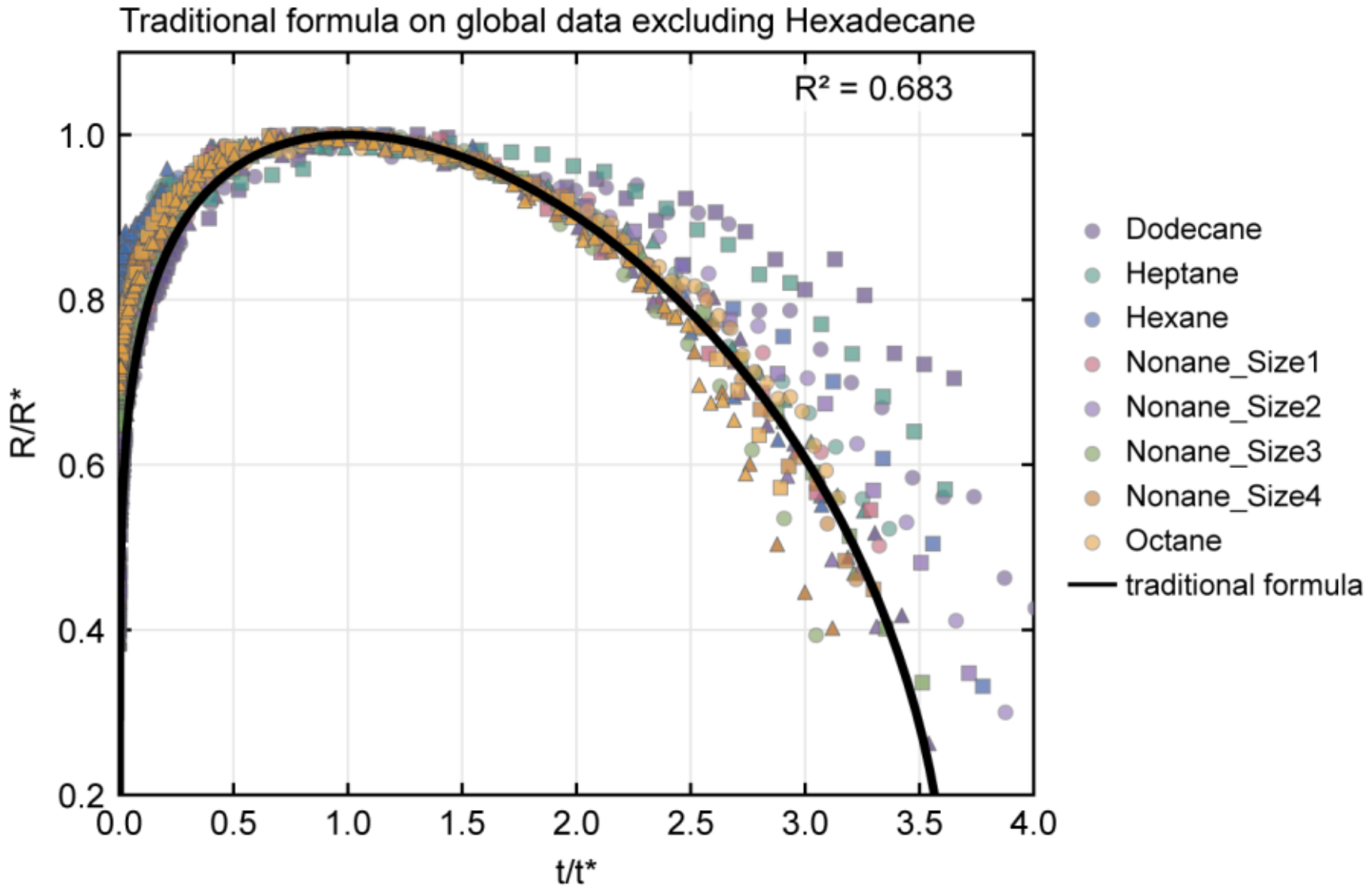

**Fig. S3. Global fit of the empirical wetting formula.** Normalized wetting trajectories across liquids and droplet-size conditions. Points show experimental data, and the black line shows the traditional formula ($R^2$=0.683).

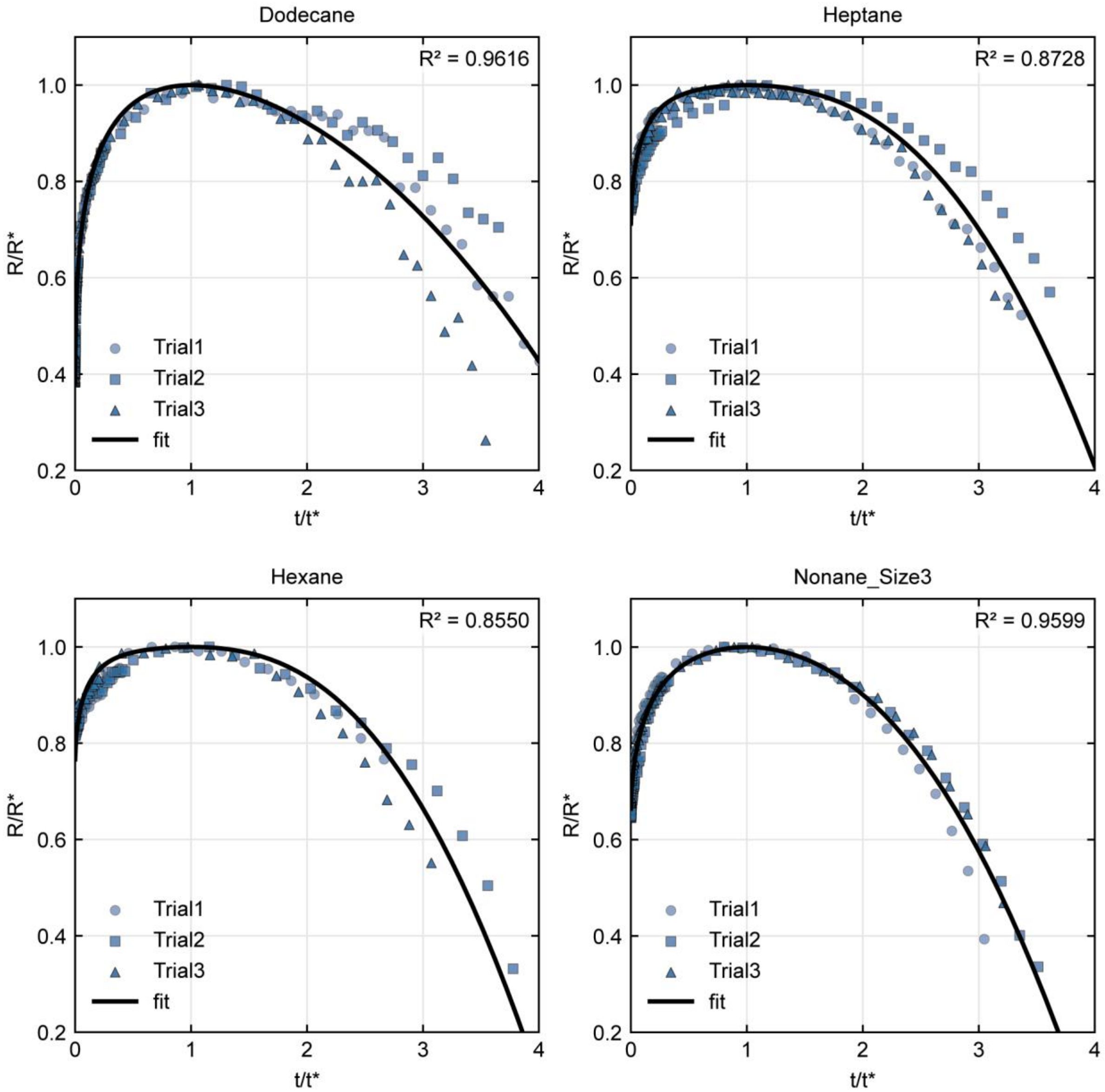

**Fig. S4. Fits to representative wetting trajectories with discovered closed physical equation.** Normalized trajectories for dodecane, heptane, hexane and the third nonane size condition. Symbols distinguish three experimental trials, and black lines show the fitted curves.

**Table S1. Sample sizes and train-test splits used in the problems for comparison.** $N$ denotes the total number of samples in the dataset.

| Case | $N$ | Inputs | Outputs | Train (80%) | Test (20%) |
|---|---|---|---|---|---|
| Keyhole | 90 | 7 | 1 | 72 | 18 |
| Dixit | 144 | 6 | 1 | 115 | 29 |
| Rayleigh–Bénard | 182 | 7 | 1 | 145 | 37 |
| MHD generator | 483 | 6 | 1 | 386 | 97 |
| Blasius | 10,000 | 4 | 1 | 8,000 | 2,000 |
| Colebrook | 100,000 | 5 | 1 | 80,000 | 20,000 |
| Rayleigh | 100,000 | 4 | 1 | 80,000 | 20,000 |

**Table S2. Generated basic dimensionless groups and optimized exponent vector $\beta$ for seven problems tested in this study.**

| Case | Generated basic Dimensionless groups | Optimized $\beta$ |
|---|---|---|
| Keyhole | $\Pi_1 = \frac{V_s r_0}{\alpha}, \Pi_2 = \frac{\eta P}{{V_s}^3 \rho (r_0)^2}, \Pi_3 = \frac{{V_s}^2}{c_p \Delta T}$ | [-1/2, -1, -1] |
| Dixit | $\Pi_1 = \frac{U_\infty \delta}{\nu}, \Pi_2 = \frac{U_\infty \delta^*}{\nu}, \Pi_3 = \frac{U_\infty \theta}{\nu}, \Pi_4 = \frac{{U_\infty}^2 \delta}{M}$ | [0, 1, -1/2, 1] |
| Rayleigh–Bénard | $\Pi_1 = \Delta T \alpha, \Pi_2 = \frac{h^3 \Delta T g \alpha}{\nu \kappa}, \Pi_3 = \frac{\nu}{\kappa}$ | [0, 1, 0] |
| MHD generator | $\Pi_1 = \frac{l^3 \rho \, \mathrm{d}p/\mathrm{d}x}{\mu^2}, \Pi_2 = \frac{\mu B^6}{\rho^2 \eta^3 (\mathrm{d}p/\mathrm{d}x)^2}$ | [1, 0] |
| Blasius | $\Pi_1 = \frac{U x}{\nu}, \Pi_2 = \frac{U y}{\nu}$ | [-1/2, 1] |
| Colebrook | $\Pi_1 = \frac{u \rho D}{\mu}, \Pi_2 = \frac{u \rho k}{\mu}$ | [1, -1] |
| Rayleigh | $\Pi_1 = \frac{U y}{\nu}, \Pi_2 = \frac{y^2}{t \nu}$ | [0, 1] |